\documentclass[11pt]{article}
\usepackage{amssymb}
\usepackage{amsmath}
\newcommand{\equal}{\!\!\!&=&\!\!\!}

\begin{document}
\abovedisplayshortskip 12pt
\belowdisplayshortskip 12pt
\abovedisplayskip 12pt
\belowdisplayskip 12pt
\baselineskip=15pt
\title{{\bf Super Extensions of the  Short Pulse Equation}}
\author{J. C. Brunelli\thanks{\texttt{jcbrunelli@gmail.com}}  \\
\\
Departamento de F\'\i sica, CFM\\
Universidade Federal de Santa Catarina\\
Campus Universit\'{a}rio, Trindade, C.P. 476\\
CEP 88040-900\\
Florian\'{o}polis, SC, Brazil\\
}
\date{}
\maketitle

\begin{center}
{ \bf Abstract}
\end{center}

From a super extension of the Wadati, Konno and Ichikawa scheme for integrable systems and using a $\mathrm{osp(1,2)}$ valued connection 1-form we obtain  super generalizations for the Short Pulse equation as well for the Elastic Beam equation.
\bigskip

\noindent {\it PACS:} 02.30.Ik; 02.30.Jr; 05.45.-a

\noindent {\it Keywords:} Short pulse equation; Integrable models; Nonlinear evolution equations;

\newpage

\section{Introduction}

Super and supersymmetric extensions of integrable systems have been investigated for a long time.  Extensive motivations and the interesting properties that have been established from these studies can be found in the literature and references therein. We cite the classical two-dimensional spacetime field theoretic models such as the supersymmetric sine-Gordon system \cite{Girardello1978}, supersymmetric Liouville model \cite{Chaichian1978} and supersymmetric $\sigma$ models \cite{D'Auria1980}. We also have the classical integrable systems such as the supersymmetric Toda lattices \cite{Ol'shanetsky1983}, super extensions of the Korteweg-de Vries (KdV) equation \cite{Kupershmidt1984}, supersymmetric extensions of the KdV equation \cite{Manin1985,Mathieu1988}, supersymmetric Kadomtsev-Petviashvili (KP) hierarchy \cite{Manin1985}, super extensions of the  Nonlinear Schr\"{o}dinger (NLS) equation \cite{Kulish1985},  supersymmetric NLS equation \cite{Roelofs1992,Brunelli1995}, supersymmetric Two Boson equation \cite{Brunelli1994}, among many others.

The main purpose of this paper is to show that for the integrable Short Pulse (SP) equation 
\begin{equation}
u_{xt} =u +{1\over 6}\left(u^3\right)_{xx}\quad{\rm or}\quad u_{t} =\left(\partial^{-1}u\right) +{1\over 2}u^2u_x\;,\label{SP}
\end{equation}
a super extension does exist. The SP equation  originally appeared in differential geometry from the study of pseudospherical surfaces \cite{Rabello1989}. Later it appeared in nonlinear optics in the study of the propagation of ultra short optical pulses in nolinear media where its integrability was numerically supported  \cite{Schafer2004,Chung2005}. Sakovich and Sakovich \cite{Sakovich2005} studied the integrability of the SP equation from a zero curvature (ZC) point of view. For the linear spectral problem
\begin{equation}
\Psi_x=\mathbb{A}\Psi\;,\quad \Psi_t=\mathbb{B}\Psi\,,\label{eigenvalue}
\end{equation}
the corresponding ZC representation
\begin{equation}
\mathbb{A}_t-\mathbb{B}_x+\left[\mathbb{A},\mathbb{B}\right]=0\;,\label{zc}
\end{equation}
is given by
\begin{equation}
\mathbb{A}=\left(
                   \begin{array}{cc}
                     \lambda & \lambda u_x\\\noalign{\vskip 10pt}
                     \lambda u_x & -\lambda\\
                   \end{array}
                 \right)\;,\qquad
                 \mathbb{B}=\left(
                   \begin{array}{cc}
                     {\lambda\over2}u^2+{1\over 4\lambda} & {\lambda\over6}\left(u^3\right)_{x}- {1\over2}u\\\noalign{\vskip 10pt}
                      {\lambda\over6}\left(u^3\right)_{x}+ {1\over2}u& -{\lambda\over2}u^2-{1\over 4\lambda} \\
                   \end{array}
                 \right)\;.\label{zcsp}
\end{equation}
In the same Ref. \cite{Sakovich2005} a transformation relating the SP to the sine-Gordon equation was given. In \cite{Sakovich2006} this transformation was used to derive exact solutions of the SP equation from known soliton solutions of the sine-Gordon equation. The recursion operator, Hamiltonian structures, conservation laws and the SP hierarchy were studied in \cite{Brunelli2005} and \cite{Brunelli2006}. Hirota's bilinear representation \cite{Kuetche2007}, multisoliton solutions \cite{Matsuno2007}, and periodic solutions \cite{Parkes2008,Matsuno2008}, among others properties, were also investigated.

Generalizations of the SP equation were studied in \cite{Sakovich2016}-\cite{Hone2016}, but in order to incorporate effects of polarization and anisotropy two-component $(u,v)$ integrable generalizations of the SP equation were proposed \cite{Pietrzyk2008}-\cite{YaoZeng2011}. These systems reduce to the SP equation (\ref{SP}) if $u=0$ or $v=0$ while others systems if $u=v$. The integrability of these systems were obtained mainly from a ZC representation or bilinear formalism and in \cite{Brunelli2013,Brunelli2012} from a Hamiltonian point of view. Let us consider a complex generalization of the SP equation
\begin{equation}
u_{xt} =u +{1\over 2}\left(uu^*u_x\right)_{x}\;,\label{complexSP}
\end{equation}
its integrability was established in \cite{Feng2015}. We can write (\ref{complexSP}) in the equivalent form
\begin{eqnarray}
u_{xt} \equal u +{1\over 2}\left(uvu_x\right)_{x}\;,\notag\\
v_{xt} \equal v +{1\over 2}\left(uvv_x\right)_{x}\;,\label{matsuno}
\end{eqnarray}
where $v\equiv u^*$. Equation (\ref{matsuno}) has the formn of the  Dimakis--M{\"{u}}ller-Hoissen--Matsuno system for a two-component SP equations for independent variable $u$ and $v$ which is also known to be integrable \cite{Dimakis2010,Matsuno2011}
 with ZC given by
\begin{equation}
\mathbb{A}=\left(
                   \begin{array}{cc}
                     \lambda & \lambda u_x\\\noalign{\vskip 10pt}
                     \lambda v_x & -\lambda\\
                   \end{array}
                 \right)\;,\qquad
                 \mathbb{B}=\left(
                   \begin{array}{cc}
                     {\lambda\over2}uv+{1\over 4\lambda} & {\lambda\over2}uvu_x- {1\over2}u\\\noalign{\vskip 10pt}
                     {\lambda\over2}uvv_x + {1\over2}v& -{\lambda\over2}uv-{1\over 4\lambda} \\
                   \end{array}
                 \right)\;.\label{zcmatsuno}
\end{equation}
Also, observe that we get two SP equations when $u=v$.

In \cite{Brunelli2005} we have shown that the equation
 \begin{equation}
 u_t=\left[{u_{xx}\over (1+u_x^2)^{3/2}}\right]_x\label{eb}
\end{equation}
is a negative flow of the SP hierarchy. The $x$ derivative of this equation is related to the Elastic Beam (EB) equation derived in \cite{WKIa,WKIb}, so we simply call (\ref{eb}) the EB  equation. Its ZC is given by
\begin{equation*}
\mathbb{A}=\left(
                   \begin{array}{cc}
                     \lambda & \lambda u_x\\\noalign{\vskip 10pt}
                     \lambda u_x & -\lambda\\
                   \end{array}
                 \right)\;,\qquad
                 \mathbb{B}=\left(
                   \begin{array}{cc}
                     {4\Delta^{-1/2}\lambda^3} & \Box_{x}\lambda+2\Box\lambda^2+4u_x\Delta^{-1/2}\lambda^3\\\noalign{\vskip 10pt}
                    \Box_{x}\lambda-2\Box\lambda^2+4u_x\Delta^{-1/2}\lambda^3& {-4\Delta^{-1/2}\lambda^3} \\
                   \end{array}
                 \right)\;,\label{zcelasticbeam}
\end{equation*}
where $\Delta\equiv 1+u_x^2$ and $\Box\equiv u_{xx}\Delta^{-3/2}$. We can easily generalize this system as a two-component EB equations
\begin{eqnarray}
u_{t} \equal \left[{u_{xx}\over (1+u_xv_x)^{3/2}}\right]_{x}\;,\notag\\
\noalign{\vskip 5pt}
v_{t} \equal \left[{v_{xx}\over (1+u_xv_x)^{3/2}}\right]_{x}\;,\label{2eb}
\end{eqnarray}
with ZC
\begin{eqnarray}
\mathbb{A}\equal\left(
                   \begin{array}{cc}
                     \lambda & \lambda v_x\\\noalign{\vskip 10pt}
                     \lambda u_x & -\lambda\\
                   \end{array}
                 \right)\;,\qquad\notag\\
 \noalign{\vskip 5pt}
                 \mathbb{B}\equal\left(
                   \begin{array}{cc}
                     (v_x\Box^u-u_x\Box^v)\lambda^2+{4\Delta^{-1/2}\lambda^3} &
                      \Box^v_{x}\lambda+2\Box^v\lambda^2+4v_x\Delta^{-1/2}\lambda^3\\\noalign{\vskip 10pt}
                    \Box^u_{x}\lambda-2\Box^u\lambda^2+4u_x\Delta^{-1/2}\lambda^3&-(v_x\Box^u-u_x\Box^v)\lambda^2-{4\Delta^{-1/2}\lambda^3} \\
                   \end{array}
                 \right)\!\;,\label{zc2eb}
\end{eqnarray}
where $\Delta\equiv 1+u_xv_x$, $\Box^u\equiv u_{xx}\Delta^{-3/2}$ and $\Box^v\equiv v_{xx}\Delta^{-3/2}$.

This paper is organized as follows. In Section \ref{WKI} we review the super extension of the Wadati, Konno and Ichikawa scheme for integrable systems. A class of super integrable equations is obtained for a $\mathrm{osp(1,2)}$-valued connection via ZC. We then apply these equations, and solve them in detail, to obtain in Sections \ref{superEB} and \ref{superSP} the super extensions of the EB and SP equations. In Section \ref{conclusions} we make some final remarks and comments.

\section{$\mathbf{osp(1,2)}$ WKI Equations}\label{WKI} 

As it is well known  completely integrable evolution equations can be written as the solution of consistency conditions for overdetermined systems of linear partial-differential equations. Geometrically, the linear system (\ref{eigenvalue}) can be written as
\begin{equation*}
d\Psi-\Omega\Psi=0\;,
\end{equation*}
where $d$ is the exterior derivative, the connection $\Omega$ is a matrix one-form,
\begin{equation*}
\Omega=\mathbb{A}\,dx+\mathbb{B}\,dt\;,
\end{equation*}
and $\Psi$ is the Jost wavefunction. The integrability of this system, the ZC (\ref{zc}), implies the condition
\begin{equation}
d\Omega-\Omega\wedge\Omega=0\;,\label{1formzc}
\end{equation}
which states that $\Omega$ is a flat connection 1-form.
By suitably parametrizing the connection $\Omega$
we obtain examples of integrable nonlinear equations. In the AKNS scheme \cite{AKNS} we consider a connection valued in $\mathrm{sl}(2,\mathbb{R})$ algebra \cite{Crampin1977,Gurses1981} and we derive the KdV equation, the modified mKdV equation, the NLS equation and the sine-Gordon equation. Wadati, Konno and Ichikawa (WKI) \cite{WKIa,WKIb}, still using a connection valued in $\mathrm{sl}(2,\mathbb{R})$ algebra but with a different parametrization, found a new series of integrable nonlinear evolution equations. In fact, as acknowledged by Sakovich \cite{Sakovich2005} the ZC formulation (\ref{zcsp}), as well (\ref{zcmatsuno}) and (\ref{zc2eb}), is of the WKI type.

 A super extension for the AKNS method was proposed by G\"urses and Oguz \cite{Gurses1985} (and revisited by Mathieu and Thibeault in \cite{Mathieu1989})  considering a connection valued in the super Lie algebra $\mathrm{osp}(N,2)$. Among general class of coupled nonlinear evolution equations they obtained for $N=1$ the well known integrable fermionic extensions of the KdV \cite{Kupershmidt1984} and NLS \cite {Kulish1985} equations. Erbay and Ogus proposed a super extension of the WKI scheme \cite{Erbay1985}. As observed by Popowicz \cite {Popowicz1990} these geometric super generalizations do  not yield supersymmetric equations since supersymmetry invariance is broken, therefore in this paper we are restrict to fermionic generalizations of the equations under investigation.

In order to derive the super extensions of the SP and EB equations we will follow the work of Erbay and Oguz. We start first embedding the $\mathrm{sl}(2,\mathbb{R})$ algebra into a super algebra, the $\mathrm{osp}(1,2)$ algebra. We use a $3\times 3$ representation defined by three bosonic generators $e_i\;, i=0,1,2$ and two fermionic generators $q_a\;, a=1,2\/$ given by
\begin{eqnarray*}
&&e_0=\left(
                   \begin{array}{rrr}
                     1 & 0 & 0 \\\noalign{\vskip 5pt}
                     0 & -1 & 0\\\noalign{\vskip 5pt}
                     0 &0&0\\
                   \end{array}
                 \right)\;,\quad
 e_1=\left(
                   \begin{array}{rrr}
                     0 & 1 & 0 \\\noalign{\vskip 5pt}
                     0 & 0 & 0\\\noalign{\vskip 5pt}
                     0 &0&0\\
                   \end{array}
                 \right)\;,\quad
 e_2=\left(
                   \begin{array}{rrr}
                     0 & 0 & 0 \\\noalign{\vskip 5pt}
                     1 & 0 & 0\\\noalign{\vskip 5pt}
                     0 &0&0\\
                   \end{array}
                 \right)\;,\quad
                 \nonumber\\\nonumber\\\nonumber\\
&&q_1=\left(
                   \begin{array}{rrr}
                     0 & 0 & 1 \\\noalign{\vskip 5pt}
                     0 & 0 & 0\\\noalign{\vskip 5pt}
                     0 &-1&0\\
                   \end{array}
                 \right)\;,\quad
 q_2=\left(
                   \begin{array}{rrr}
                     0 & 0 & 0 \\\noalign{\vskip 5pt}
                     0 & 0 & 1\\\noalign{\vskip 5pt}
                     1 &0&0\\
                   \end{array}
                 \right)\;.
\end{eqnarray*}
The generators have commutation, $[\ ,\ ]$ and anticommutation, $[\ ,\ ]_+$, relations given by
\begin{equation*}
\begin{array}{lll}
{[e_0,e_1]=2e_1}\quad & {[e_0,e_2]=-2e_2}\quad & {[e_1,e_2]=e_0} \\
\noalign{\vskip 5pt}
{[e_0,q_1] = q_1} & {[e_0,q_2]=-q_2} & {[e_1,q_1]=0}  \\
\noalign{\vskip 5pt}
{[e_1,q_2]=q_1} & {[e_2,q_1]=q_2} & {[e_2,q_2]=0}   \\
\noalign{\vskip 5pt}
{[q_1,q_2]_+=e_0} & {[q_1,q_1]_+=-2e_1} & {[q_2,q_2]_+=2e_2}\;. \\
\end{array}
\end{equation*}
A super soliton connection 1-form $\omega$ with values in  the $\mathrm{osp}(1,2)$ algebra is defined by
\begin{equation}
\Omega=e_i\theta_i+q_a\pi_a=\left(\begin{array}{rrr}
\theta_0 &\theta_1 &\pi_1\\
\theta_2 &-\theta_0 & \pi_2\\
\pi_2 & -\pi_1 &0\\
\end{array}\right)\;,\label{1formconnection}
\end{equation}
and we use the following parametrization for the 1-forms $\theta_i$ and $\pi_a$
\begin{eqnarray}
&&\theta_0 = A(t,x,\lambda)\,dt+\lambda\, p(t,x)\,dx\;, \notag \\\noalign{\vskip 5pt}
&&\theta_1 = C(t,x,\lambda)\,dt+\lambda\, r(t,x)\,dx\;, \notag \\\noalign{\vskip 5pt}
&&\theta_2 =  B(t,x,\lambda)\,dt+\lambda\, q(t,x)\,dx\;, \notag \\\noalign{\vskip 5pt}
&&\pi_1 =  \alpha(t,x,\lambda)\,dt+\lambda\,\beta(t,x)\,dx\;, \notag \\\noalign{\vskip 5pt}
&&\pi_2 =  \rho(t,x,\lambda)\,dt+\lambda\,\epsilon(t,x)\,dx\;,\label{parametrization}
\end{eqnarray}
yielding the decomposition
\begin{equation}
\Omega=\underbrace{\left(\begin{array}{rrr}
\lambda p &\lambda r &\lambda\beta\\
\lambda q&-\lambda p & \lambda\epsilon\\
\lambda\epsilon & -\lambda\beta &0\\
\end{array}\right)}_{\displaystyle=\mathbb{A}}dx+
\underbrace{\left(\begin{array}{rrr}
A &C &\alpha\\
B&-A & \rho\\
\rho & -\alpha &0\\
\end{array}\right)}_{\displaystyle=\mathbb{B}}dt
\;.\label{abdecomposition}
\end{equation}
In (\ref{parametrization}) $A$, $B$, $C$ are commuting, $\alpha$ and $\rho$ are anticommuting functions while $p$, $r$, $q$ are commuting, and $\beta$ and $\epsilon$ are anticommuting potentials.

Substituting (\ref{1formconnection}) with (\ref{parametrization}) in the ZC condition (\ref{1formzc}) we obtain the equations
\begin{eqnarray*}
&&A_x+\lambda\left(-p_t+qC-rB+\alpha\epsilon-\beta\rho\right)=0\;, \notag \\\noalign{\vskip 5pt}
&&B_x+\lambda\left(-q_t-2qA+2pB-2\epsilon\rho\right)=0\;, \notag \\\noalign{\vskip 5pt}
&&C_x+\lambda\left(-r_t-2pC+2rA-2\alpha\beta\right)=0\;, \notag \\\noalign{\vskip 5pt}
&&\alpha_x+\lambda\left(-\beta_t+A\beta -p\alpha+C\epsilon-r\rho\right))=0\;, \notag \\\noalign{\vskip 5pt}
&&\rho_x+\lambda\left(-\epsilon_t+B\beta-q\alpha-A\epsilon +p\rho\right)=0\;.
\label{eqZC}
\end{eqnarray*}
From these equations we can get super integrable nonlinear partial differential evolution equations. We substitute the power series of the spectral parameter $\lambda$
\begin{eqnarray}
A=\sum_{n=-M}^N a_n\lambda^n\;,\quad B=\sum_{n=-M}^N b_n\lambda^n\;,\quad C=\sum_{n=-M}^Nc_n\lambda^n\;,  \notag\\
\noalign{\vskip 6pt}
\alpha=\sum_{n=-M}^N \alpha_n\lambda^n\;,\quad \rho=\sum_{n=-M}^N \rho_n\lambda^n\;,\quad\quad\quad\quad\quad\label{zcseries}
\end{eqnarray}
and we obtain, after comparing powers of $\lambda$, the following recursion relations for $n\ge-(M-1)$ and $n\not=1$
\begin{eqnarray}
&&a_{n,x}+qc_{n-1}-rb_{n-1}-\epsilon\alpha_{n-1}-\beta\rho_{n-1}=0\;, \label{arec} \\\noalign{\vskip 5pt}
&&b_{n,x}-2qa_{n-1}+2pb_{n-1}-2\epsilon\rho_{n-1}=0\;, \label{brec} \\\noalign{\vskip 5pt}
&&c_{n,x}-2pc_{n-1}+2ra_{n-1}+2\beta\alpha_{n-1}=0\;, \label{crec} \\\noalign{\vskip 5pt}
&&\alpha_{n,x}+\beta a_{n-1}-p\alpha_{n-1}+\epsilon c_{n-1}-r\rho_{n-1}=0\;, \label{alpharec}\\\noalign{\vskip 5pt}
&&\rho_{n,x}+\beta b_{n-1}-q\alpha_{n-1}-\epsilon a_{n-1}+p\rho_{n-1}=0\;,
\label{rhorec}
\end{eqnarray}
and the nonlinear evolution equations for $n=1$
\begin{eqnarray}
p_t\equal a_{1,x}+qc_0-rb_0-\epsilon\alpha_0-\beta\rho_0\;, \label{peq} \\\noalign{\vskip 5pt}
q_t\equal b_{1,x}-2qa_0+2pb_0-2\epsilon\rho_0 \;, \label{qeq} \\\noalign{\vskip 5pt}
r_t\equal  c_{1,x}-2pc_0+2ra_0+2\beta\alpha_0\;, \label{req} \\\noalign{\vskip 5pt}
\beta_t\equal \alpha_{1,x}+\beta a_0-p\alpha_0+\epsilon c_0-r\rho_0 \;, \label{betaeq} \\\noalign{\vskip 5pt}
\epsilon_t\equal \rho_{1,x} +\beta b_0-q\alpha_0-\epsilon a_0+p\rho_0\;,
\label{epsiloneq}
\end{eqnarray}
where $a_{-M}$, $b_{-M}$, $c_{-M}$, $\alpha_{-M}$ and $\rho_{-M}$ are constants. Now we will consider two specific cases for $M$ and $N$ in (\ref{zcseries}):

\section{Elastic Beam Super Extension: $M=-1$ and $N=3$}\label{superEB}

We start with $n=4$ and we set
\[
a_4=b_4=c_4=\alpha_4=\rho_4=0\;.
\]
Equations (\ref{brec})--(\ref{rhorec}) yields the linear system
\[
{\renewcommand\arraystretch{2}
\left(\begin{array}{c|c}
\mathbf A & \mathbf B\\
\hline
\mathbf C & \mathbf D\\
\end{array}\right)}
\left(\begin{array}{r}
b_3\\
c_3\\
\hline
\alpha_3\\
\rho_3\\
\end{array}\right)=
\left(\begin{array}{r}
2qa_3\\
-2ra_3\\
\hline
-\beta a_3\\
\epsilon a_3\\
\end{array}\right)\;.
\]
The even supersymmetric matrix
\begin{equation}
\mathbb{M}=\left(\begin{array}{cc}
\mathbf A & \mathbf B\\
\mathbf C & \mathbf D\\
\end{array}\right)\;,\label{evensusymatrix}
\end{equation}
with
\[
{\mathbf A}=\left(\begin{array}{cc}
2p & 0\\
0  & -2p\\
\end{array}\right)\;,\quad
{\mathbf B}=\left(\begin{array}{cc}
0 & -2\epsilon\\
2\beta  & 0\\
\end{array}\right)\;,\quad
{\mathbf C}=\left(\begin{array}{cc}
0 & \epsilon\\
\beta  & 0\\
\end{array}\right)\;,\quad
{\mathbf D}=\left(\begin{array}{cc}
-p & -r\\
-q  & p\\
\end{array}\right)\;,\quad
\]
has the inverse
\[
\mathbb{M}^{-1}=\left(\begin{array}{cc}
({\mathbf A}-{\mathbf B}{\mathbf D}^{-1}{\mathbf C})^{-1} & -{\mathbf A}^{-1}{\mathbf B}({\mathbf D}-{\mathbf C}{\mathbf A}^{-1}{\mathbf B})^{-1}\\
\noalign{\vskip 15pt}
-{\mathbf D}^{-1}{\mathbf C}({\mathbf A}-{\mathbf B}{\mathbf D}^{-1}{\mathbf C})^{-1} & ({\mathbf D}-{\mathbf C}{\mathbf A}^{-1}{\mathbf B})^{-1}\\
\end{array}\right)\;.
\]
Therefore
\[
\mathbb{M}^{-1}=\left(\begin{array}{cccc}
\displaystyle{{1\over 2p}{\Delta\over \Delta+\epsilon\beta}} & \displaystyle{0} & \displaystyle{-{{q\epsilon/p}\over \Delta-2\epsilon\beta}} & \displaystyle{\epsilon\over \Delta-2\epsilon\beta} \\
\noalign{\vskip 15pt}
\displaystyle{0} & \displaystyle{-{1\over 2p}{\Delta\over \Delta+\epsilon\beta}}  & \displaystyle{-{{\beta}\over \Delta-2\epsilon\beta}} & \displaystyle{-{{r\beta/p}\over \Delta-2\epsilon\beta}} \\
\noalign{\vskip 15pt}
\displaystyle{{1\over 2}{{r\beta/p}\over \Delta+\epsilon\beta}} & \displaystyle{-{1\over 2}{\epsilon\over \Delta+\epsilon\beta}} & \displaystyle{{(\epsilon\beta-p^2)/p}\over \Delta-2\epsilon\beta} & \displaystyle{-{r\over \Delta-2\epsilon\beta}} \\
\noalign{\vskip 15pt}
\displaystyle{-{1\over 2}{\beta\over \Delta+\epsilon\beta}} & \displaystyle{-{1\over 2}{q\epsilon/p\over \Delta+\epsilon\beta}} & \displaystyle{-{q\over \Delta-2\epsilon\beta}} & \displaystyle{-{{(\epsilon\beta-p^2)/p}\over \Delta-2\epsilon\beta}} \\
\end{array}\right)\;,
\]
where $\Delta=p^2+qr$. Then we have as solution
\[
b_3={q\over p}a_3\;,\quad c_3={r\over p}a_3\;,\quad \alpha_3= { \beta\over p }a_3\;,\quad \rho_3={ \epsilon \over p }a_3\;,
\]
where $a_3$ is an arbitrary function. In this way (\ref{arec}) is identically satisfied.

For $n=3$, equation (\ref{arec}) with (\ref{brec})--(\ref{rhorec}) for $b_2$, $c_2$, $\rho_2$ and $\alpha_2$ gives the ode
\[
\left[(p^2+rq-2\epsilon\beta)a_3^2\right]_x=0\;,
\]
with solution
\[
a_3=C^0\,\Delta^{-1/2}+C^0\,\epsilon\beta\Delta^{-3/2}\;,
\]
where $C^0$ is a constant. Returning to (\ref{brec})--(\ref{rhorec}) we obtain the linear system
\[
\mathbb{M}
\left(\begin{array}{r}
b_2\\
c_2\\
\alpha_2\\
\rho_2\\
\end{array}\right)=
\left(\begin{array}{r}
-b_{3,x}+2qa_2\\
-c_{3,x}-2ra_2\\
-\alpha_{3,x}-\beta a_2\\
-\rho_{3,x}+\epsilon a_2\\
\end{array}\right)\;,
\]
with solution
\begin{eqnarray}
b_2\equal -{1\over2p}\,b_{3,x} +{1\over p}\,qa_2+\left({1\over2p}\,\epsilon\beta b_{3,x}+{1\over p}\,q\epsilon\alpha_{3,x}-\epsilon\rho_{3,x}\right)\Delta^{-1}\;, \notag \\\noalign{\vskip 5pt}
c_2\equal {1\over2p}\,c_{3,x} +{1\over p}\,ra_2+\left(-{1\over2p}\,\epsilon\beta c_{3,x}+{1\over p}\,r\beta\rho_{3,x}+\beta\alpha_{3,x}\right)\Delta^{-1}\;, \notag \\\noalign{\vskip 5pt}
\alpha_2\equal {1\over p}\,\beta a_2+\left(-{1\over2p}\,r\beta b_{3,x}+{1\over 2}\,\epsilon c_{3,x}+p\alpha_{3,x}-{1\over p}\,\epsilon\beta\alpha_{3,x}+r\rho_{3,x}\right)\Delta^{-1}+ \notag \\\noalign{\vskip 5pt}
&\!\!\!\! \!\!\!\!&\qquad +\left(2p\epsilon\beta\alpha_{3,x}+2r\epsilon\beta\rho_{3,x}\right)\Delta^{-2}\;,\notag \\\noalign{\vskip 5pt}
\rho_2\equal {1\over p}\,\epsilon a_2+\left({1\over2p}\,q\epsilon c_{3,x}+{1\over 2}\,\beta b_{3,x}-p\rho_{3,x}+{1\over p}\,\epsilon\beta\rho_{3,x}+q\alpha_{3,x}\right)\Delta^{-1}- \notag \\\noalign{\vskip 5pt}
&\!\!\!\! \!\!\!\!&\qquad -\left(2p\epsilon\beta\rho_{3,x}-2q\epsilon\beta\alpha_{3,x}\right)\Delta^{-2}\;,\label{n2solutions}
\end{eqnarray}
where we use the ansatz for the unknown $a_2$
\begin{equation}
a_2=f+g\epsilon\epsilon_x+h\beta\beta_x+\ell\beta\epsilon+m\beta_x\epsilon+n\beta\epsilon_x\;,\label{a2ansatz}
\end{equation}
with $f$, $g$, $h$, $\ell$, $m$ and $n$ to be determined. Analogously, for $n=2$ we get from (\ref{brec})--(\ref{rhorec})
\begin{eqnarray*}
b_1\equal -{1\over2p}\,b_{2,x} +{1\over p}\,qa_1+\left({1\over2p}\,\epsilon\beta b_{2,x}+{1\over p}\,q\epsilon\alpha_{2,x}-\epsilon\rho_{2,x}\right)\Delta^{-1}\;, \notag \\\noalign{\vskip 5pt}
c_1\equal {1\over2p}\,c_{2,x} +{1\over p}\,ra_1+\left(-{1\over2p}\,\epsilon\beta c_{2,x}+{1\over p}\,r\beta\rho_{2,x}+\beta\alpha_{2,x}\right)\Delta^{-1}\;, \notag \\\noalign{\vskip 5pt}
\alpha_1\equal {1\over p}\,\beta a_1+\left(-{1\over2p}\,r\beta b_{2,x}+{1\over 2}\,\epsilon c_{2,x}+p\alpha_{2,x}-{1\over p}\,\epsilon\beta\alpha_{2,x}+r\rho_{2,x}\right)\Delta^{-1}+ \notag \\\noalign{\vskip 5pt}
&\!\!\!\! \!\!\!\!&\qquad +\left(2p\epsilon\beta\alpha_{2,x}+2r\epsilon\beta\rho_{2,x}\right)\Delta^{-2}\;,\notag \\\noalign{\vskip 5pt}
\rho_1\equal {1\over p}\,\epsilon a_1+\left({1\over2p}\,q\epsilon c_{2,x}+{1\over 2}\,\beta b_{2,x}-p\rho_{2,x}+{1\over p}\,\epsilon\beta\rho_{2,x}+q\alpha_{2,x}\right)\Delta^{-1}- \notag \\\noalign{\vskip 5pt}
&\!\!\!\! \!\!\!\!&\qquad -\left(2p\epsilon\beta\rho_{2,x}-2q\epsilon\beta\alpha_{2,x}\right)\Delta^{-2}\;.\label{n1solutions}
\end{eqnarray*}
We set $a_1=0$ and solving (\ref{arec}) for $a_2$ given by (\ref{a2ansatz}) we obtain
\begin{equation*}
a_2=\left[{1\over 4p}\,C^0\,\Delta(rq_x-qr_x)-{1\over 2p}\,(2p^2-rq)(\epsilon_x\beta-\epsilon\beta_x)+{3\over2}\,r\epsilon\epsilon_x-{3\over2}\,q\beta\beta_x\right]\Delta^{-5/2}\;,
\end{equation*}
and we can finally write (\ref{n2solutions}) as
\begin{eqnarray*}
b_2\equal {1\over2}\,C^0\left\{-C^1-{1\over p}\,\left[(2p^2-rq)\epsilon\epsilon_x+3q^2\beta\beta_x\right]\Delta^{-5/2}\right\}\;, \notag \\\noalign{\vskip 5pt}
c_2\equal {1\over2}\,C^0\left\{C^2+{1\over p}\,\left[(2p^2-rq)\beta\beta_x+3r^2\epsilon\epsilon_x\right]\Delta^{-5/2}\right\}\;, \notag \\\noalign{\vskip 5pt}
\alpha_2\equal {1\over 2}\,C^0\left(\Pi^1+{1\over p}\,\Sigma^2\right)\;, \notag \\\noalign{\vskip 5pt}
\rho_2\equal {1\over 2}\,C^0\left({1\over p}\,\Pi^2-\Sigma^1\right)\;,\label{n2finalsolutions}
\end{eqnarray*}
where
\begin{eqnarray*}
C^1\equal q_x\Delta^{-3/2}+3q(\epsilon_x\beta-\epsilon\beta_x)\Delta^{-5/2}\;, \notag \\\noalign{\vskip 5pt}
C^2\equal r_x\Delta^{-3/2}-3r(\epsilon_x\beta-\epsilon\beta_x)\Delta^{-5/2}\;, \notag \\\noalign{\vskip 5pt}
\Pi^1\equal \left(\beta\Delta^{-3/2}\right)_x+\beta_x\Delta^{-3/2} \;, \notag \\\noalign{\vskip 5pt}
\Pi^2\equal \left(q\beta\Delta^{-3/2}\right)_x+q\beta_x\Delta^{-3/2} \;, \notag \\\noalign{\vskip 5pt}
\Sigma^1\equal \left(\epsilon\Delta^{-3/2}\right)_x+\epsilon_x\Delta^{-3/2} \;, \notag \\\noalign{\vskip 5pt}
\Sigma^2\equal \left(r\epsilon\Delta^{-3/2}\right)_x+r\epsilon_x\Delta^{-3/2}\;.\label{placeholders}
\end{eqnarray*}
For $n=0$  we obtain from (\ref{arec})--(\ref{rhorec}) that
\[
a_0=b_0=c_0=\alpha_0=\rho_0=0\;.
\]

We consider that $p$ is constant and therefore equation (\ref{peq}) is satisfied and from (\ref{qeq})--(\ref{epsiloneq}) we obtain  the following  evolution equations
\begin{eqnarray}
q_t\equal {1\over 4p}\, C^0\left\{C^1_x+{1\over p}\,\left(\left[(2p^2-rq)\epsilon\epsilon_x+3q^2\beta\beta_x\right]\Delta^{-5/2}\right)_x-\epsilon\beta\Delta^{-1}\left(q_x\Delta^{-3/2}\right)_x\right.+ \notag \\\noalign{\vskip 5pt}
&\!\!\!\! \!\!\!\!&\qquad+\left.2q\epsilon\Delta^{-1}\left(\Pi^1+{1\over p}\Sigma^2\right)_x-2p\epsilon\Delta^{-1}\left({1\over p}\Pi^2-\Sigma^1\right)_x\right\}_x\;,\notag \\\noalign{\vskip 5pt}
r_t\equal {1\over 4p}\, C^0\left\{C^2_x+{1\over p}\,\left(\left[(2p^2-rq)\beta\beta_x+3r^2\epsilon\epsilon_x\right]\Delta^{-5/2}\right)_x-\epsilon\beta\Delta^{-1}\left(r_x\Delta^{-3/2}\right)_x\right.+ \notag \\\noalign{\vskip 5pt}
&\!\!\!\! \!\!\!\!&\qquad+\left.2r\beta\Delta^{-1}\left({1\over p}\Pi^2-\Sigma^1\right)_x+2p\beta\Delta^{-1}\left(\Pi^1+{1\over p}\Sigma^2\right)_x\right\}_x\;,\notag \\\noalign{\vskip 5pt}
\beta_t\equal {1\over 4p}\, C^0\left\{ r\beta\Delta^{-1}\left[C^1+{1\over p}\,(2p^2-rq)\epsilon\epsilon_x\Delta^{-5/2}\right]_x+p\epsilon\Delta^{-1}\left[C^2+{1\over p}\,(2p^2-rq)\beta\beta_x\Delta^{-5/2}\right]_x\right.+ \notag \\\noalign{\vskip 5pt}
&\!\!\!\! \!\!\!\!&\!\!\!\! \!\!+\left.2p\left(p-{1\over p}\,\epsilon\beta+2p\epsilon\beta\Delta^{-1}\right)\Delta^{-1}\left(\Pi^1+{1\over p}\Sigma^2\right)_x+2pr\left(1+2\epsilon\beta\Delta^{-1}\right)\Delta^{-1}\left({1\over p}\Pi^2-\Sigma^1\right)_x
\right\}_x\;,\notag \\\noalign{\vskip 5pt}
\epsilon_t\equal {1\over 4p}\, C^0\left\{ q\epsilon\Delta^{-1}\left[C^2+{1\over p}\,(2p^2-rq)\beta\beta_x\Delta^{-5/2}\right]_x-p\beta\Delta^{-1}\left[C^1+{1\over p}\,(2p^2-rq)\epsilon\epsilon_x\Delta^{-5/2}\right]_x\right.- \notag \\\noalign{\vskip 5pt}
&\!\!\!\! \!\!\!\!&\!\!\!\! \!\!-\left.2p\left(p-{1\over p}\,\epsilon\beta+2p\epsilon\beta\Delta^{-1}\right)\Delta^{-1}\left({1\over p}\Pi^2-\Sigma^1\right)_x+2pq\left(1+2\epsilon\beta\Delta^{-1}\right)\Delta^{-1}\left(\Pi^1+{1\over p}\Sigma^2\right)_x
\right\}_x\;.\notag \\\label{qrebeq}
\end{eqnarray}
These equations are in agreement with the ones reported in \cite{Erbay1985} (equations (7) in that paper). Setting $C^0=4$, $p=1$, $\epsilon=\beta=0$, $q=u_x$ and $r=v_x$ we obtain the two-component EB equations (\ref{2eb}) as well the ZC (\ref{zc2eb}) from $\mathbb{A}$ and $\mathbb{B}$ in (\ref{abdecomposition}).

Now we perform the reduction $q=u_x$, $r=k_1\overline{u}_x$, $\epsilon=\psi_x$ and $\beta=k_2\overline{\psi}_x$, where $k_1$ and $k_2$ are constants and a bar over a quantity denotes the Berezin conjugation in the Grassman algebra such that $\overline{\psi\phi}=\overline{\phi}\overline{\psi}$. In the case of usual complex valued functions, this operation reduces to the usual complex conjugation \cite{Gurses1985,Erbay1985}. Therefore, equations (\ref{qrebeq}) reduce to the following super EB equations
\begin{eqnarray}
u_t\equal C^1_x+\left(\left[(2-k_1|u_x|^2)\psi_{x}\psi_{xx}+3k_2^2u_x^2\overline{\psi}_{x}\overline{\psi}_{xx}\right]\Delta^{-5/2}\right)_x-k_2\psi_{x}\overline{\psi}_{x}\Delta^{-1}\left(u_{xx}\Delta^{-3/2}\right)_x + \notag \\\noalign{\vskip 5pt}
&\!\!\!\!+ \!\!\!\!&\!\!\!  2u_x\psi_{x}\Delta^{-1}\left(k_2\,\overline{\Sigma}^1+\Sigma^2\right)_x+{2\over k_2}\psi_{x}\Delta^{-1}\left(k_2\,{\Sigma}^1+\overline{\Sigma}^2\right)_x \;,\notag \\\noalign{\vskip 10pt}
\psi_t\equal  k_1u_x\psi_{x}\Delta^{-1}\left[\overline{C}^1-(2-k_1|u_x|^2)\overline{\psi}_{x}\overline{\psi}_{xx}\Delta^{-5/2}\right]_x-\notag \\\noalign{\vskip 5pt}
&\!\!\!\!- \!\!\!\!&\!\!\!k_2\overline{\psi}_{x}\Delta^{-1}\left[C^1+(2-k_1|u_x|^2)\psi_{x}\psi_{xx}\Delta^{-5/2}\right]_x + \notag \\\noalign{\vskip 5pt}
&\!\!\!\!+ \!\!\!\!&\!\!\! {2\over k_2}\left(1-k_2\psi_{x}\overline{\psi}_{x}+2k_2\psi_{x}\overline{\psi}_{x}\Delta^{-1}\right)\Delta^{-1}\left(k_2\,\Sigma^1+\overline{\Sigma}^2\right)_x+\notag \\\noalign{\vskip 10pt}
&\!\!\!\!+ \!\!\!\!&\!\!\! 2u_x\left(1+2k_2\psi_{x}\overline{\psi}_{x}\Delta^{-1}\right)\Delta^{-1}\left(k_2\,\overline{\Sigma}^1+\Sigma^2\right)_x
 \;,\notag
\end{eqnarray}
provided that $k_1=-k_2^2$ with $k_1$ and $k_2$ real ($\overline{k}_i=k_i$).

\section{Short Pulse Super Extension: $M=1$ and $N=1$}\label{superSP}

For $n=-1$ we set
\[
 a_{-1}=\hbox{constant}\quad \hbox{and}\quad  b_{-1}=c_{-1}=\alpha_{-1}=\rho_{-1}=0\;.
\]
For $n=0$ equations (\ref{arec})--(\ref{rhorec}) give
\begin{eqnarray*}
a_{0}\equal \hbox{constant}\;, \notag \\\noalign{\vskip 5pt}
b_{0}\equal 2a_{-1}(\partial^{-1}q) \;, \notag \\\noalign{\vskip 5pt}
c_{0}\equal  -2a_{-1}(\partial^{-1}r) \;, \notag \\\noalign{\vskip 5pt}
\alpha_{0}\equal -a_{-1}(\partial^{-1}\beta)  \;, \notag \\\noalign{\vskip 5pt}
\rho_{0}\equal a_{-1}(\partial^{-1}\epsilon) \;. \notag \\\noalign{\vskip 5pt}
\end{eqnarray*}
Setting
\[
 a_{2}=b_{2}=c_{2}=\alpha_{2}=\rho_{2}=0\;,
\]
equations (\ref{brec})--(\ref{rhorec}) for $n=2$ give
\[
\mathbb{M}
\left(\begin{array}{r}
b_1\\
c_1\\
\alpha_1\\
\rho_1\\
\end{array}\right)=
\left(\begin{array}{r}
2qa_1\\
-2ra_1\\
-\beta a_1\\
\epsilon a_1\\
\end{array}\right)\;,
\]
where $\mathbb{M}$ is given by (\ref{evensusymatrix}). From the calculations in the last section we have as solutions
\[
b_1={q\over p}a_1\;,\quad c_1={r\over p}a_1\;,\quad \alpha_1= { \beta\over p }a_1\;,\quad \rho_1={ \epsilon \over p }a_1\;.
\]
Here $a_1$ is an arbitrary function and equation (\ref{arec}) is identically satisfied.

To get local equations from (\ref{qeq})--(\ref{epsiloneq}) we set $q=Q_x$, $r=R_x$, $\beta=\Phi_x$, $\epsilon=\Psi_x$ to obtain
\begin{eqnarray}
p_t\equal \left(a_1-2a_{-1}QR-a_{-1}\Phi\Psi\right)_x\;, \label{sppeq} \\\noalign{\vskip 5pt}
Q_{xt}\equal \left({Q_x\over p}a_1\right)_x-2a_0Q_x+4a_{-1}p\,Q-2a_{-1}\Psi_x\Psi\;, \notag \\\noalign{\vskip 5pt}
R_{xt}\equal \left({R_x\over p}a_1\right)_x+2a_0R_x+4a_{-1}p\,R-2a_{-1}\Phi_x\Phi\;, \notag \\\noalign{\vskip -10pt}\label{finalequations}\\
\Phi_{xt}\equal \left({ \Phi_x \over p }a_1\right)_x +a_0\Phi_x+a_{-1}p\,\Phi-2a_{-1}R\Psi_x-a_{-1}R_x\Psi\;, \notag \\\noalign{\vskip 5pt}
\Psi_{xt}\equal \left({ \Psi_x \over p }a_1\right)_x -a_0\Psi_x+a_{-1}p\,\Psi+2a_{-1}Q\Phi_x+a_{-1}Q_x\Phi\;. \notag
\end{eqnarray}
For $p=1$, $Q=u$, $R=v$, $\Phi=\Psi=0$, $a_{-1}=1/4$ and  $a_0=0$ the first equation (\ref{sppeq}) gives $a_1=uv/2$ and we obtain the Dimakis--M{\"{u}}ller-Hoissen--Matsuno system (\ref{matsuno}) from the equations in (\ref{finalequations}). For $p=1+u_xv_x$, $Q=R=u-v$, $\Phi=\Psi=0$, $a_{-1}=1/4$, $a_0=0$ and choosing $a_1=(u^2+v^2)(1+u_xv_x)/2$ we obtain the Feng system of equations \cite{Feng2012}
\begin{eqnarray}
u_{xt} \equal u +uu_x^2+{1\over 2}\left(u^2+v^2\right)u_{xx}\;, \notag\\
v_{xt} \equal v +vv_x^2+{1\over 2}\left(u^2+v^2\right)v_{xx}\;,\label{eqFeng}
\end{eqnarray}
with ZC (\ref{zc}) following from $\mathbb{A}$ and $\mathbb{B}$ in (\ref{abdecomposition}),
\begin{eqnarray}
&&\mathbb{A}=\left(
                   \begin{array}{cc}
                     \lambda(1+u_xv_x) & \lambda(u_x-v_x)\\\noalign{\vskip 10pt}
                     \lambda(u_x-v_x) & -\lambda(1+u_xv_x)\\
                   \end{array}
                 \right)\;,\nonumber\\\noalign{\vskip 10pt}
              &&   \mathbb{B}=\left(
                   \begin{array}{cc}
                     {\lambda\over2}(u^2+v^2)(1+u_xv_x)+{1\over 4\lambda} & {\lambda\over2}(u^2+v^2)(u_x-v_x)- {1\over2}(u-v)\\\noalign{\vskip 10pt}
                     {\lambda\over2}(u^2+v^2)(u_x-v_x)+ {1\over2}(u-v) & -{\lambda\over2}(u^2+v^2)(1+u_xv_x)-{1\over 4\lambda} \\
                   \end{array}
                 \right)\;,\label{ZCeqFeng}
\end{eqnarray}
introduced in \cite{Brunelli2013}.

Let us now write the super SP integrable nonlinear partial differential equations following from equations (\ref{finalequations}). We set $Q=u$, $R=v$, $\Phi=\phi$, $\Psi=\psi$, $a_{-1}=1/4$ and  $a_0=0$. For $p=1$ the equation (\ref{sppeq}) gives
\[
a_1={1\over 2}\,uv+{1\over 4}\,\phi\psi\;,
\]
and we obtain
\begin{eqnarray}
u_{xt}\equal u+{1\over2}\left(uvu_x\right)_x +{1\over4}\left(\phi\psi u_x\right)_x-{1\over2}\psi_x\psi\;, \notag \\\noalign{\vskip 5pt}
v_{xt}\equal v+ {1\over2}\left(uvv_x\right)_x +{1\over4}\left(\phi\psi v_x\right)_x-{1\over2}\phi_x\phi\;, \notag \\\noalign{\vskip 5pt}
\phi_{xt}\equal {1\over 4}\phi+{1\over2}\left(uv\phi_x\right)_x+{1\over4}\left(\phi\psi\phi_x\right)_x
 -{1\over 2}v\psi_x-{1\over4}v_x\psi\;, \notag \\\noalign{\vskip 5pt}
\psi_{xt}\equal  {1\over 4}\psi+{1\over2}\left(uv\psi_x\right)_x+{1\over4}\left(\phi\psi\psi_x\right)_x
 +{1\over 2}u\phi_x+{1\over4}u_x\phi\;,\label{uvspeq}
\end{eqnarray}
with ZC 
\begin{eqnarray}
&&\mathbb{A}=\left(
                   \begin{array}{ccc}
                     \lambda & \lambda v_x & \lambda  \phi_x\\\noalign{\vskip 10pt}
                     \lambda u_x & -\lambda & \lambda \psi_x\\\noalign{\vskip 10pt}
                     \lambda \psi_x & -\lambda \phi_x & 0\notag\\
                   \end{array}
                 \right)\;,\nonumber\\\noalign{\vskip 10pt}
              &&   \mathbb{B}=\left(
                   \begin{array}{ccc}
                       {1\over4\lambda}+{1\over 4}(2uv+\phi\psi)\lambda  & -{1\over2}v+{1\over 4}v_x(2uv+\phi\psi)\lambda &  -{1\over4}\phi+{1\over 4}\phi_x(2uv+\phi\psi)\lambda\\\noalign{\vskip 10pt}
                        {1\over2}u+{1\over 4}u_x(2uv+\phi\psi)\lambda & -{1\over4\lambda}-{1\over 4}(2uv+\phi\psi)\lambda& {1\over4}\psi+{1\over 4}\psi_x(2uv+\phi\psi)\lambda \\\noalign{\vskip 10pt}
                       {1\over4}\psi+{1\over 4}\psi_x(2uv+\phi\psi)\lambda  & {1\over4}\phi-{1\over 4}\phi_x(2uv+\phi\psi)\lambda& 0\notag \\
                   \end{array}
                 \right)\;.\\\label{ZCuvspeq}
\end{eqnarray}

Again, we perform the reduction $v=k_1\overline{u}$ and $\phi=k_2\overline{\psi}$, where $k_1$ and $k_2$ are constants. Therefore, equations (\ref{uvspeq})
reduce to the following super SP equations
\begin{eqnarray}
u_{xt}\equal u+{1\over2}\,k_1\left(u\overline{u}u_x\right)_x-{1\over4}\,k_2\left(\psi\overline{\psi} u_x\right)_x-{1\over2}\psi_x\psi\;, \notag \\\noalign{\vskip 5pt}
\psi_{xt}\equal  {1\over 4}\psi+{1\over2}\,k_1\left(u\overline{u}\psi_x\right)_x-{1\over4}\,k_2\left(\psi\overline{\psi}\psi_x\right)_x
 +{1\over 2}\,k_2u\overline{\psi}_x+{1\over4}\,k_2u_x\overline{\psi}\;, \notag
\end{eqnarray}
provided that $k_1=-k_2^2$ with $k_1$ and $k_2$ real ($\overline{k}_i=k_i$).
\section{Conclusions}\label{conclusions}

In this paper, we have obtained the super extensions of the EB equation (\ref{2eb}), equation (\ref{qrebeq}) with $q=u_x$, $r=v_x$, $\beta=\phi_x$ and $\epsilon=\psi_x$, and the super extension of the SP equation (\ref{matsuno}) given by (\ref{uvspeq}) with ZC (\ref{ZCuvspeq}). As a byproduct of the bosonic part we obtained the ZC (\ref{ZCeqFeng}) for the Feng system (\ref{eqFeng}) used in \cite{Brunelli2013}. The Hamiltonian structure of these equations are under investigation.

As we have already pointed out, we performed a fermionic extension of the EB and SP equations that preserve its complete integrability, since we generalized the WKI formulation through a super Lie algebra. However, these fermionic extensions are not supersymmetric invariant, i.e., there is no transformation relating the bosonic and fermionic fields leaving the system invariant. Using superspace and superfield techniques a susy SP equation is also under investigation and will be reported elsewhere. 

\end{document}